%% file: Paper.tex
\documentclass[conference]{IEEEtran}
\input{Preamble}

\begin{document}

\title{Satellite-Aided Entanglement Distribution for Optimized Quantum Networks}

\author{\IEEEauthorblockN{Jakob Kaltoft Søndergaard}
\IEEEauthorblockA{\textit{Department of Electronic Systems} \\
\textit{Aalborg University}\\
Aalborg, Denmark \\
jakobks@es.aau.dk}
\and
\IEEEauthorblockN{René Bødker Christensen}
\IEEEauthorblockA{\textit{Department of Mathematical Sciences} \\
\textit{Aalborg University}\\
Aalborg, Denmark \\
rene@math.aau.dk}
\and
\IEEEauthorblockN{Petar Popovski}
\IEEEauthorblockA{\textit{Department of Electronic Systems} \\
\textit{Aalborg University}\\
Aalborg, Denmark \\
petarp@es.aau.dk}
}

\maketitle

\begin{abstract}
  Quantum internet needs to ensure timely provision of entangled qubits to be used in tasks that involve distributed quantum computing or sensing. This has been addressed by a top-down approach of optimized quantum networks \cite{b1}, in which entanglement is distributed prior to receiving tasks. Upon the task arrival, the desired entanglement state is attained with local operations and classical communication. The pre-distribution of entanglements should aim to minimize the amount of qubits used, as this decreases the risk of decoherence and thus degradation of the entangled state.
  The optimized quantum networks consider a multi-hop optical network and in this work we are supplementing it with \emph{Satellite-aided Entanglement Distribution (SED)}. The motivation is that satellites can shortcut the topology and place the entanglement at two nodes not directly connected through the optical network. We devise an algorithm for strategic placement of entanglements with SED, which results in a decrease in the number of qubits used in pre-distribution of entanglement. The numerical results show that SED can significantly enhance the performance of small quantum networks, while \emph{Entanglement-sharing Constraints (EC)} are critical for large networks.
\end{abstract}


\section{Introduction}
Entanglement distribution in quantum networks is essential to realize the immense potential of quantum technologies \cite{b7}, e.g., secure networking by means of quantum key distribution (QKD) \cite{b6}, unprecedented computational power via distributed quantum computing \cite{b8}, and optimized sensor technology through quantum sensing \cite{b9}. Entanglement can be distributed by physically transmitting quantum states through quantum channels, typically fiber-optic cables or terrestrial free-space to support optic-based quantum communication \cite{b11}. However, the inherent photon loss in optical communication that scales exponentially with distance \cite{b12}; the no-cloning theorem \cite{b13}, which prevents copying of quantum information; and decoherence together impose a constraint of a few hundred kilometers on terrestrial optical communication \cite{b16}. A promising approach to realize a global quantum internet is to utilize satellite technology. Even a single satellite can cover multiple locations on Earth separated by thousands of kilometers, and experimental research into the field of satellite-based entanglement distribution has successfully demonstrated distribution of entanglement between quantum devices located more than 1,200 kilometers apart with the low Earth orbit (LEO) satellite Micius \cite{b5}.

A prominent alternative to quantum networks solely relying on physical distribution of entanglement is repeater-based quantum networks, in which quantum repeaters are placed throughout the network. In this setup, entanglement is physically transmitted to nearby repeaters, which produce the requested end-to-end entanglement to the network through an appropriate sequence of entanglement swappings \cite{b14} and rounds of entanglement distillation \cite{b15} to achieve sufficiently high fidelity.
One of the key obstacles to realizing a fully functioning repeater-based quantum network is the difficulty of distributing entanglement when factors such as decoherence, throughput, latency, and limited resources at the repeaters are taken into account \cite{b17,b18,b19,b20,b21,b22,b23}. Common for most of such works is that they consider bottom-up approaches where entanglement is distributed upon task arrival, which introduces inevitable latency for two reasons. First, the optimal routing solution, which includes path selection for entanglement swapping, resource allocation decision, protocol for entanglement distillation, etc., is dynamical and must therefore be determined at every time instance the network is active. Second, the inherent randomness in entanglement generation, entanglement swapping, and entanglement distillation, generally enforces several rounds of these operations to be performed in order to obtain the end-to-end entanglement with sufficiently high fidelity.

\begin{figure}
    \centering
    \input{Figures/Phases_Time_Diagram2}
    \caption{Time-diagram of the network's phase for three network uses. The initialization phase is used after the second network use due to a change in the network. The $i$-th task arrival is at time $t_i$. The coherence times for qubits in the $i$-th network use must be lower than $\tau_i$. To reduce the risk of decoherence, the satellite can distribute entanglement at any time in the adaptive phase.}
    \label{fig: Time_Diagram_Phases}
\end{figure}
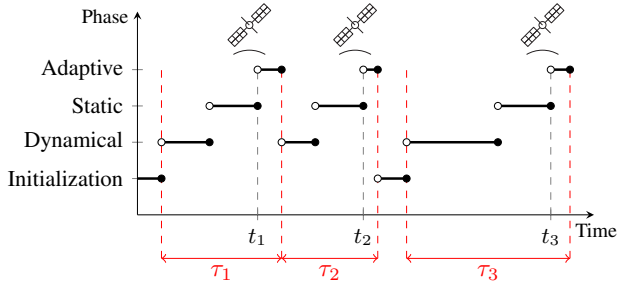

A framework with a top-down approach was recently proposed to reduce the latency that is inherent in bottom-up quantum networks \cite{b2}, \cite{b24}. In \cite{b24}, a top-down network use is partitioned into three phases, namely the dynamical, static, and adaptive phases, as illustrated in Fig.~\ref{fig: Time_Diagram_Phases}. During the dynamical phase, entanglement is generated and distributed (with quantum routing protocols) to the network according to some predetermined graph state known as the resource state. Notice that the network topology of the resource state is an abstraction of the physical network topology. After the resource state has been distributed, the static phase begins, where the network stores the resource state while sitting idle until the next task arrival. Decoherence imposes a timing constraint on the lifetime of the qubits in the resource state, including the resource state sitting idle in the static phase, as illustrated by the red arrows in Fig.~\ref{fig: Time_Diagram_Phases}. After the task arrival, the network enters the adaptive phase where the resource state is modified with local operations and classical communication (LOCC) to satisfy the received task. As only LOCC is required to satisfy a task upon task arrival in this framework, it offers a significant latency reduction compared to the time-consuming bottom-up approach relying on quantum routing. Furthermore, notice that only a single initialization round is required to determine the optimal resource state whenever a change in the network occurs as illustrated in Fig.~\ref{fig: Time_Diagram_Phases}.

In the top-down framework, different algorithms for minimizing the number of qubits in the resource state in an ideal setting were proposed in \cite{b1}, the effect of noise in simple network topologies in \cite{b3}, \cite{b33}, while algorithms for satisfying tasks while maintaining as much of the pre-distributed entanglement as possible were analyzed in \cite{b10}. In all these works, the resource state is assumed to always have been distributed to the network prior to a task arrival. This is, however, implausible in practice for two reasons. First, the limited coherence time of the resource state constraints the time that the system can be in the static phase. Second, the randomness of entanglement distribution makes it a challenging problem to determine when to begin the dynamical phase while guaranteeing it to successfully distribute the resource state prior to the next task arrival. Said differently, it is challenging to guarantee the dynamical phase to finish before the next task arrival, while also ensuring the static phase to be shorter than the coherence times of the network devices. The difficulty of this is largely dependent on the physical network topology.

In this work, we therefore propose a setting for the top-down approach where the physical network topology is taken into account. We do this in two different ways as an answer to the two above-mentioned challenges. First, to decrease the need for stored resources and thereby risk of decoherence, we enable satellite-aided entanglement distribution (SED) in the adaptive phase to complement the resource state. Second, to reduce the complexity of the quantum routing in the dynamical phase and thus the time it takes to complete, we introduce entanglement-sharing constraints (EC) between physically distant devices. Due to the minimal resources expected in quantum devices in the near future, our goal is to minimize the stored resources in the resource state, i.e., number of qubits, during the static phase similarly to \cite{b1}. In this paper, we therefore address the following question:\medskip

\textit{How is the timeliness of optimized quantum network affected by satellite-aided entanglement distribution and entanglement constraints, and how do these concepts influence the need for stored resources in optimized resource states?}\medskip

The contribution of this work can be summarized as:
\begin{itemize}
    \item We propose a realistic setting for the top-down approach by introducing EC in the dynamical phase and SED in the adaptive phase.
    \item We generalize the merging algorithm from \cite{b1} to reduce the number of stored resources in our proposed setting.
    \item We make numerical simulations of the performance of networks when considering/not considering SED and EC.
\end{itemize}

The remainder of this paper is structured as follows. In Sec. \ref{sec: Preliminaries}, we introduce preliminaries about graph states and graph-related matrices. We present our network model in Sec. \ref{sec: Model}. In Sec. \ref{sec: Optimizing_Resource} we introduce an algorithm for determining an optimized resource state in our setting, which is the foundation for the numerical results presented in Sec. \ref{sec: Results}. Lastly, we conclude in Sec. \ref{sec: Conclusion}.

\section{Preliminaries}\label{sec: Preliminaries}
\subsection{Graph states}
Graph states \cite{b4} are a particular class of quantum states that can be characterized by graphs. More precisely, the graph state $\ket{G}$ associated with the undirected graph $G=(V,E)$ is defined as
\begin{equation}
    \ket{G}=\left(\prod_{(a,b)\in E} CZ_{a,b}\right)\ket{+}^{\otimes\abs{V}},
\end{equation}
where $CZ_{a,b}$ is the controlled phase operator with control $a$ and target $b$. From the definition, it follows that the vertices in $G$ correspond to qubits in $\ket{G}$ while edges indicate an entangling operation between the incident vertices. Graph states are often more conveniently defined in the stabilizer formalism \cite{b29} as the unique $+1$-eigenstate of the stabilizers
\begin{equation}
    S_v = X_v\prod_{u\in N(v)}Z_u,
\end{equation}
for all $v\in V$, where $N(v)$ is the neighborhood of $v$ and $X_a$ ($Z_a$) is the Pauli $X$ ($Z$) operator on qubit $a$. Although a graph state is uniquely determined by its corresponding graph, it is often more useful to consider the set of states that is local unitary equivalent to it, i.e., the states satisfying $\ket{G'}=U\ket{G}$ for some local unitary operator $U$.

\textit{Example:} Let $G$ be the simple graph with an edge connecting its two vertices $a$ and $b$. By definition, 
\begin{equation}
    \ket{G}=\frac{1}{2}\bigg[\ket{00}+\ket{01}+\ket{10}-\ket{11}\bigg].
\end{equation}
Applying a Hadamard operator to either of the qubits yields the EPR pair
\begin{equation}
    \ket{G'}=\frac{1}{\sqrt{2}}[\ket{00}+\ket{11}].
\end{equation}

Graph states are directly related to the topology of its corresponding graph up to a local unitary operation. This implies that graph states can describe multipartite entangled systems but also a system consisting of several smaller separate entangled subsystems. For example, a graph where each vertex has degree one corresponds to a system consisting of separate bipartite EPR pairs.

Measurements of qubits in a graph state are also elegantly described in terms of operations in graph theory. Up to local correction operations depending on the measurement outcome, the graph operations for some common measurements include:
\begin{enumerate}
    \item A $Z$-measurement of qubit $v$ corresponds to removing vertex $v$ from $G$ as well as all its incident edges.
    \item A $Y$-measurement of qubit $v$ corresponds to performing a local complementation of $G$ with respect to $v$ and then performing a $Z$-measurement of $v$.
    \item An $X$-measurement of qubit $v$ corresponds to performing a local complementation of $G$ with respect to a vertex $u$ in the neighborhood of $v$, making a $Y$-measurement of $v$, and then another local complementation with respect to $u$.
    \item A projective \emph{merging} measurement of two neighboring qubits $v$ and $u$ with observables $\{P_0,P_1\}$ defined as
    \begin{align}
        P_0 &=\ket{0}_w\bra{00}_{u,v}+\ket{1}_w\bra{11}_{u,v}\\
        P_1 &=\ket{0}_w\bra{01}_{u,v}+\ket{1}_w\bra{10}_{u,v}
        \end{align}
        corresponds to performing an edge contraction of edge ($v,u$) of $G$ yielding the merged qubit $w$ (without multiple edges) \cite{b32}.
\end{enumerate}
It should be noted that local complementations can also be performed without measurements, but solely with LOCC \cite{b4}. If a graph, $G$, can be obtained from another graph, $G'$, through a sequence of local complementations, then $G$ and $G'$ are said to be locally equivalent. Determining whether two graphs are locally equivalent is generally a NP-complete problem \cite{b26,b27}.

\subsection{Graph-related matrices}\label{sec: Graph_matrices}
For an undirected graph $G=(V,E)$, the adjacency matrix \cite{b28}, denoted $A^G$, is defined as the symmetric matrix with nonzero entries indicating the multiplicity of an edge between two vertices, i.e.,
\begin{equation}
    A^G_{i,j} = \begin{cases}
                  w(i,j), & \text{if } (i,j)\in E,\\
                  0, & \text{otherwise,}
              \end{cases} 
\end{equation}
where $w(i,j)$ is the multiplicity of edge $(i,j)$ in $E$. If $G$ is a simple graph, $w(i,j)=1$ for all $(i,j)\in E$.

For a set of graphs $\mathcal{G}=\{G^{(k)}\}_k$, we define the union adjacency matrix as
\begin{equation}
    U^\mathcal{G}_{i,j} = \max_k A^{G^{(k)}}_{i,j}.
\end{equation}
Nonzero entries in $U^{\mathcal{G}}$ indicate the maximal multiplicity of an edge between two vertices in any of the graphs in $\mathcal{G}$. Notice that the union adjacency matrix is identical to the adjacency matrix defined in \cite{b1} when $\mathcal{G}$ only contains simple graphs.

The simultaneous adjacency matrix $S^\mathcal{G}$ \cite{b1} of a set of graphs $\mathcal{G}$ contains information about the maximum number of edges in a graph containing a given edge,
i.e.,
\begin{equation}
    S^\mathcal{G}_{i,j}=\max_k \Big(A_{i,j}^{G^{(k)}}\sum_{n<m}A^{G^{(k)}}_{n,m}\Big).
\end{equation}
These matrices build the foundation for optimizing the resource state in Sec.~\ref{sec: Optimizing_Resource}.

\section{Network model}\label{sec: Model}
\begin{figure}
    \centering
    \input{Figures/Physical_Network_with_Satellite}
    \caption{Physical network topology considered in this work. The terrestrial network devices being users and quantum repeaters are represented by red squares and gray circles, respectively. The blue lines represents optical fiber cables and the satellite coverage area is the orange circle.}
    \label{fig: Physical_Network_with_Satellite}
\end{figure}
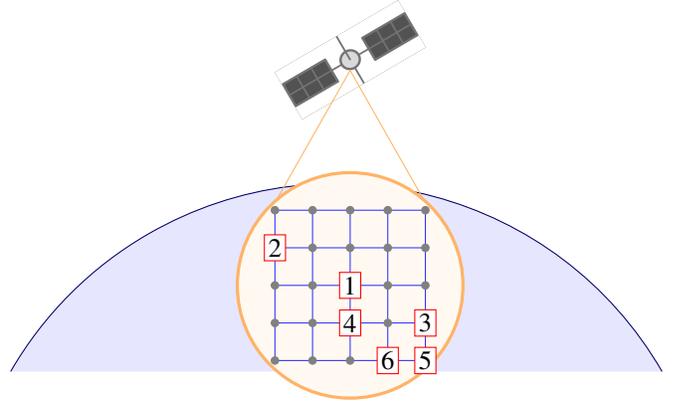

In this work, we want to establish entanglement between users in a hypothetical but realistic wide area network (WAN). For simplicity, we consider a homogeneous terrestrial network with an equidistant $5\times5$ grid topology consisting of both users and quantum repeaters as depicted in Fig.~\ref{fig: Physical_Network_with_Satellite}. We assume that devices are connected by optical fiber of length $200$ km such that the maximal distance between any pair of devices is approximately $1,131$ km. To supplement the terrestrial network, we introduce a LEO satellite with the entire terrestrial network within its coverage.

To establish end-to-end entanglement between distant users in the network, we use the top-down approach where pre-distributed entanglement in the resource state (dynamical phase) can be modified by LOCC (adaptive phase) upon task arrival.

\textbf{Definition:} $R$ denotes the chosen resource state, i.e., the entanglement that has been pre-distributed to the network prior to task arrivals. 

In this work, we focus on the initialization phase by determining optimized resource states. Hence, we do not explicitly consider the distribution of the resource state, however, its complexity is motivation for the introduction of SED and EC.

\subsection{Entanglement-sharing constraints on resource states}\label{sec: Physical_constraints}
Guaranteeing that the resource state has been successfully distributed with high fidelity prior to a task arrival is extremely challenging, which constraints the applicability of the top-down approach where this is a general assumption. This is particularly true for large physical networks. To address this, we introduce EC on the resource state that can be distributed during the dynamical phase. This is done by constraining the maximal rounds of entanglement swapping that can be performed in each dynamical phase by a distance threshold we denote $D$. Equivalently, this constraints which users that can share an entangling operation in the resource state, i.e., an edge in the graph representation of $R$. Notice that the distance counts the number of devices between the two end users. Hence, if $D=0$, only users connected by a direct channel may share an entangling operation in the resource state, while any pair of users may share one if $D$ is larger than the maximal distance in the network, which is $D=7$ in Fig.~\ref{fig: Physical_Network_with_Satellite}. Although two distant users cannot be directly connected by an edge in the resource state, they can be indirectly entangled through an intermediate user with distance less than $D$ from both users if such exists. The end-to-end entangling operation can be obtained if the intermediate user shares an EPR pair with each of the end users whereafter a merging measurement is performed on the two qubits at the intermediate user. Hence the intermediate user effectively acts as a quantum repeater for the end users. This generalizes naturally to longer paths with several intermediate users.

\subsection{Satellite-aided entanglement distribution in the adaptive phase}\label{sec: satellite_distribution}
Another simplification of the top-down approach is that the entanglement distribution is limited to the dynamical phase due to the latency that would otherwise be introduced. However, this completely removes the capability to physically distribute entanglement on demand between devices that are connected by a quantum channel with generally negligible latency. In this work, we therefore modify the top-down approach such that entanglement also can be physically distributed between devices that are connected by channels. More precisely, we consider a simplistic model where we assume that the terrestrial network always is in the coverage of a LEO satellite as depicted in Fig.~\ref{fig: Physical_Network_with_Satellite}, such that the satellite can distribute entanglement (SED) between any pair of users on demand. We assume that SED is always successful and has negligible latency compared to the latency occurring in the adaptive phase when performing LOCC to modify the resource state to the task due to the higher propagation speed in free space. To make this more realistic, we furthermore assume that the satellite can only provide a single EPR pair in each adaptive phase.

\subsection{Network tasks}
Many different types of entanglement may be requested in a quantum network including EPR pairs for quantum teleportation, QKD, and distributed quantum computing; maximally entangled multipartite states for quantum secret sharing \cite{b30} or quantum metrology \cite{b31}; and more generally, non-maximally entangled multipartite states for blind quantum computing \cite{b34}. For simplicity, network tasks considered in this work are assumed to only consist of either a single or multiple separate EPR pairs based on current knowledge of the network.

\textbf{Definition:} $T^{(k)}$ denotes the $k$-th possible network task and can be represented as a simple graph where each node has degree at most one. Each edge in the graph represents a requested EPR pair.

The network is therefore capable of directly satisfying bipartite applications. Due to the infinitely many possible tasks that may arrive simultaneously, it is impossible to determine a reasonably sized resource state capable of satisfying all possible incoming tasks with LOCC and SED without further assumptions. We therefore adopt the usual assumption for the top-down approach, namely that only a single task from a finite set of tasks may arrive at once.

\textbf{Definition:} $\mathcal{T}=\{T^{(k)}\}_k$ with $\abs{\mathcal{T}}<\infty$ denotes a set of network tasks that may be requested by the users in the network.

Although this seemingly is a great simplification considering the dynamical nature of network tasks, the resource state to distribute in each dynamical phase can be changed along with changes in network tasks as illustrated by the two initialization phases in Fig.~\ref{fig: Time_Diagram_Phases}. This will, however, not be considered in this work. Lastly, we assume that the resource state has always been successfully distributed to the network prior to the following task arrivals. How realistic this is depends on the topology of the resource state, the distance threshold $D$, and the physical implementation of the network devices. Nonetheless, we refrain from discussing physical implementations in this work as the framework is compatible with any suitable implementation.

\subsection{Network performance}
We say that the network successfully satisfies an incoming task, $T^{(k)}$, if it is deterministically capable of modifying the resource state, $R$, to the task with LOCC and the entanglement provided by the satellite. If the network can successfully satisfy all possible tasks in a set of tasks, $\mathcal{T}$, we say that the network is successful with respect to that set of tasks. The performance of a successful network is measured as the storage requirement in the static phase, i.e., number of qubits in the resource state $R$, which we denote $Q(R)$. It should be noted that we consider an ideal setting in the sense that LOCC is assumed to be error-free and of negligible cost. Furthermore, other practically important properties such as decoherence, fidelity, link failures, and robustness against photon losses are not considered in this work for simplicity. With the setting of this work in order we can precisely define the optimization problem of this work:
\begin{alignat}{2}
    &\text{minimize } &&Q(R)\\
    &\quad\text{ s.t. } &&R\in\mathcal{R}_\mathcal{T},
\end{alignat}
where $\mathcal{R}_\mathcal{T}$ is the set of resource states that make the network successful with respect to $\mathcal{T}$ under EC, i.e., the set of resource states capable of satisfying all tasks in $\mathcal{T}$ with LOCC and SED while having no edges of length greater than $D$.

Similarly to the original setting, we need to find a resource state that is locally equivalent to a set of tasks. Checking this equivalence is still NP-complete, so we do not expect to be able to find optimal solutions in general. Instead, we aim to find a `good enough' resource state with a simple algorithm.

\subsection{Network example}\label{sec: Network_Example}
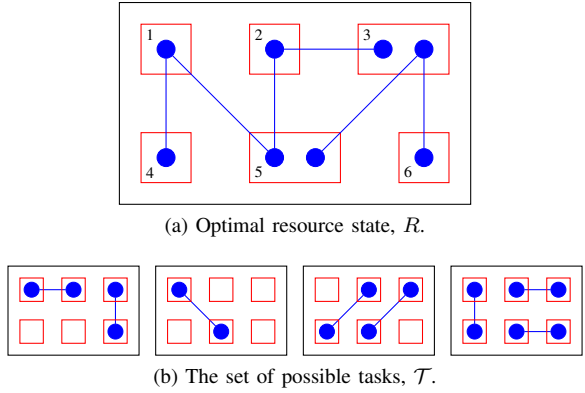
\begin{figure} 
    \centering
    \subfloat[Optimal resource state, $R$.]{%
        \input{Figures/Paper_Example_Resource}}
    \\
    \subfloat[The set of possible tasks, $\mathcal{T}$.\label{fig: Example_requests}]{%
        \input{Figures/Paper_Example_Requets}}
    \caption{The example in \cite[Fig.~1]{b1} containing the optimal resource state for the set of possible tasks. Each black square represents the abstract topology of users, red squares are users, blue circles are qubits, and blue lines indicate entanglement.}
    \label{fig: Example_from_Optimized_Quantum_Networks}
\end{figure}

To illustrate our model, consider the example provided in \cite[Fig.~1]{b1} and replicated in Fig.~\ref{fig: Example_from_Optimized_Quantum_Networks} containing the optimal resource state consisting of eight qubits for a network with six users and four possible tasks. The first task can, for example, be satisfied by performing the following sequence of measurements from left to right: $Y_{5_l}Z_{3_l}Z_4Z_{5_r}$, where $l,r$ indicates that it is on the given user's left or right qubit, respectively.

The two generalizations we consider in this work, SED and EC, are now introduced separately for the example. We assume that the positions of users in the abstract network topology in Fig.~\ref{fig: Example_from_Optimized_Quantum_Networks} coincides with the positions in the physical network topology in Fig.~\ref{fig: Physical_Network_with_Satellite}. That is, user $5$ in the abstract topology is positioned in the bottom right corner in the physical topology, and so on.

\subsubsection{Introducing EC}
\begin{figure} 
    \centering
    \subfloat[Optimal resource state when $D=5$.\label{fig: Network_example_d=5}]{%
        \input{Figures/Example_Physical_Solution_D=5}}
    \hspace{5pt}
    \subfloat[Possible resource state when $D=4$.\label{fig: Network_example_d=4}]{%
        \input{Figures/Example_Physical_Solution_D=4}}
    \caption{Possible resource states for the tasks in Fig.~\ref{fig: Example_from_Optimized_Quantum_Networks} for different choices of $D$ when considering EC.}
    \label{fig: Example_Physical_Network_Different_D}
\end{figure}
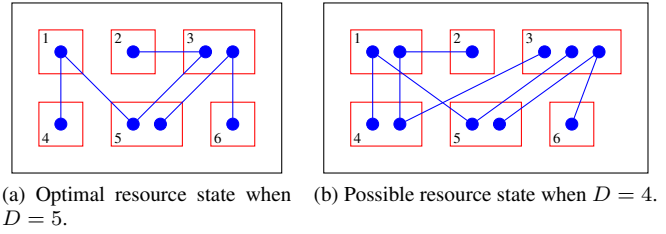

To distribute the resource state in Fig.~\ref{fig: Example_from_Optimized_Quantum_Networks}, entanglement between, for example, users $1$ and $4$ is needed, which can easily be generated with the channel connecting the users in the physical network. However, it also requires entanglement between users $2$ and $5$, which have a distance of six. Assuming that $D<6$, this introduces excessive latency due to the complexity of quantum routing, hence the resource state in Fig.~\ref{fig: Example_from_Optimized_Quantum_Networks} cannot be used as a resource state in our setting. If $D=5$ such that users $2$ and $5$ cannot share an edge but all other combinations of users can, a possible resource state with the same number of qubits can be obtained by making a local complementation of the qubit at user $2$ followed by a local complementation at the left qubit at user $3$. The resulting resource state is shown in Fig.~\ref{fig: Network_example_d=5}. For this resource state, the first task can be satisfied by the following measurements: $X_{5_l}Z_{3_l}Z_4Z_{5_r}$, where the neighbor used for the $X$-measurement is the left qubit of user $3$. In this particular example, it is possible to find another resource state that is locally equivalent to that of the original resource state while satisfying EC. This is, however, not always the case. For example, if $D=4$, user $2$ can neither share an edge with users $3, 5$, nor $6$, implying that no locally equivalent graph is capable of satisfying all tasks. Therefore, a resource state with additional qubits is required. A possible solution is shown in Fig.~\ref{fig: Network_example_d=4}, which requires $11$ qubits corresponding to a $37.5\%$ increase from the other resources states. In this case, the first task can be satisfied by performing the measurements $Z_{4_r}Z_{5_r}$ when the other entangled qubits are ignored. Notice that for $D<2$, the network is unsuccessful as user $2$ cannot be entangled with any of the other users in the network.

\subsubsection{Introducing SED}
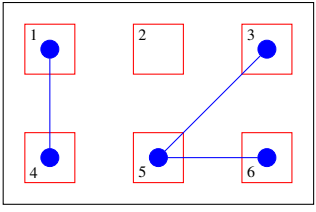
\begin{figure}
    \centering
    \input{Figures/Example_Solution3_Resource}
    \caption{Possible resource state for the set of tasks in Fig.~\ref{fig: Example_requests} when allowing SED.}
    \label{fig: Example_Satellite_Distribution}
\end{figure}

When SED is considered, it is generally possible to reduce the resource state as exemplified by the resource state in Fig.~\ref{fig: Example_Satellite_Distribution} that is capable of satisfying any of the four tasks in Fig.~\ref{fig: Example_requests} when complemented by the EPR pair provided by the satellite. To see this, let $S_{a,b}$ denote the operation of the satellite providing an EPR pair to users $a$ and $b$. When ignoring irrelevant qubits, the operations needed to satisfy each of the tasks are; 1) $S_{1,2}Y_5$, 2) $S_{1,5}$, 3) $S_{2,4}Z_6$, and 4) $S_{2,3}Z_3$. Hence, with SED, a mere five qubits suffice to satisfy all possible network tasks corresponding to a $37.5\%$ reduction compared to the original case without considering EC and SED in Fig.~\ref{fig: Example_from_Optimized_Quantum_Networks}. It should be emphasized that the two qubits needed for the satellite to distribute an EPR pair are not included in $Q(R)$ as they are not stored in the static phase.

In fact, the resource state in Fig.~\ref{fig: Example_Satellite_Distribution} can be distributed solely with channels in the physical network as only neighboring users in the physical network share an edge in the resource state. Hence, the dynamical phase would for this resource state be fairly simple as there are no EC on the resource state regardless of the chosen distance threshold $D$. This implies that the network is successful for any $D$, as opposed to being true for $D\geq2$ without SED. If $D$ is chosen such that the network is successful without SED, the performance of the network can only be enhanced by introducing SED. Comparing the resource states in Fig.~\ref{fig: Network_example_d=4} and Fig.~\ref{fig: Example_Satellite_Distribution}, which is without and with SED, respectively, for $D=4$, it can be seen that $54.4\%$ fewer qubits suffice by introducing SED. By reducing $D$, the performance enhancement is only increased. These considerations truly demonstrate the potential of  enabling SED in the adaptive phase of the top-down approach.

\section{Optimizing resource states}\label{sec: Optimizing_Resource}
In this section, we introduce algorithms for determining the resource state in our setting based on the graph-related matrices presented in Sec.~\ref{sec: Graph_matrices}. The computational complexity of the algorithms are summarized in Table~\ref{tab: Algorithm_Complexity} together with the worst-case performance of their resulting resource states. Recalling that $R$ denotes the resource state and $\mathcal{T}$ is a set of tasks, each of which can be represented as a simple graph by assumption, the interpretation of the graph-related matrices are: $A^R_{i,j}$ indicates how many edges user $i$ and $j$ share in the resource state, $U^\mathcal{T}_{i,j}$ indicates how many EPR pairs user $i$ and $j$ at most share in any of the tasks, and $S^\mathcal{T}_{i,j}$ indicates the maximal number of EPR pairs in the subset of tasks that include one between user $i$ and $j$.

\subsection{Naïve approach}
A naïve solution to determine $R$ presented in \cite{b1} is to let it correspond to $U^\mathcal{T}$. This corresponds to distributing every EPR pair that occurs in any of the task in $\mathcal{T}$ implying that the network is successful without LOCC. It can therefore be considered as a sort of worst-case solution. For the network in Fig.~\ref{fig: Example_from_Optimized_Quantum_Networks}, this would require $Q(R)=16$ qubits, as there are $8$ different EPR pairs requested in the set of tasks, corresponding to a $100\%$ increase from the optimal solution. This penalty in performance comes from not utilizing LOCC. In general, $Q(R)$ equals the number of nonzero entries in $U^\mathcal{T}$ implying that for a network of $N$ users, then
\begin{equation}\label{eq: Complexity_Naive_Solution}
    Q(R)=\begin{cases}
        \mathcal{O}(N\abs{\mathcal{T}}), & \text{if } \abs{\mathcal{T}}<N-1,\\
        \mathcal{O}(N^2), & \text{if } \abs{\mathcal{T}}\geq N-1.
    \end{cases}
\end{equation}
This is due to the assumption of each user requesting EPR pairs with at most $\min(\abs{\mathcal{T}},N-1)$ other users among the tasks. We will use this naïve solution as a benchmark for algorithms that work in our proposed setting and henceforth denote the resource state found with this approach $R_\text{BM}$.

\subsection{Satellite-aided entanglement distribution}\label{sec: Optimizing_SED}
When considering SED, the resource state can be complemented by strategically distributing an EPR pair between any two users in the network. Different approaches can be used to determine which EPR pair the satellite should provide for each of the possible tasks. In this work, we consider the case where upon task arrival, the satellite distributes one of the EPR pairs in the received task. As the resource state no longer needs to provide this EPR pair, it corresponds to removing an edge in each task. Due to the complexity of quantum routing for large networks, we let the satellite provide the EPR pair of longest length in each task (if there are several, one of these is chosen at random). Hence, the resource state when considering SED, $R_\text{SED}$, is obtained by the following steps:
\begin{enumerate}
    \item For every $k=1,\ldots,\abs{\mathcal{T}}$, let $(a^{(k)},b^{(k)})$ denote the edge of longest length in $T^{(k)}$.
    \item Let $\mathcal{T}_\text{SED}=\{T^{(k)}_\text{SED}\}_k$, where $T^{(k)}_\text{SED}=T^{(k)}\setminus\{(a^{(k)},b^{(k)})\}$.
    \item Let $R_\text{SED}$ correspond to $U^{\mathcal{T}_\text{SED}}$.
\end{enumerate}
As $T^{(k)}_\text{SED}$ is a subgraph of $T^{(k)}$ for every $k$, $R_\text{SED}$ is a subgraph of $R_\text{BM}$ implying that the performance of the network when considering SED is at least as good as the naïve approach, i.e., $Q(R_\text{SED})\leq Q(R_\text{BM})$. However, the asymptotic growth of $Q(R_\text{SED})$ is equal to that of $Q(R_\text{BM})$ in \eqref{eq: Complexity_Naive_Solution}.

\subsection{Entanglement-sharing constraints}\label{sec: Optimizing_EC}
When EC is considered, edges in $R_\text{BM}$ longer than $D$ are divided into paths such that no edge in the path has length greater than $D$, if possible. If not, the network fails to satisfy $\mathcal{T}$. In this work, we consider the simple approach of choosing the shortest path between the end users (if there are several, one of these is chosen at random). Notice that this implies that the resource state when considering EC, $R_\text{EC}$, may be a multigraph. The steps to obtain resource state $R_\text{EC}$ are:
\begin{enumerate}
    \item For each pair of users $(i,j)$ with $U^\mathcal{T}_{i,j}=1$, calculate the Manhattan distance,  $d_{i,j}$, between the users.
    \item If $d_{i,j}>D$, find the shortest path $p_{i,j}$ between $i$ and $j$ such that none of its edges have length greater than $D$.
    \item Let $\mathcal{T}_\text{EC}=\{T^{(k)}_\text{EC}\}_k$, where $T^{(k)}_\text{EC}=(T^{(k)}\setminus\{(i,j)\})\cup p_{i,j}$ if $(i,j)\in E^{(k)}$, and $T^{(k)}$ otherwise.
    \item Let $R_\text{EC}$ correspond to $U^{\mathcal{T}_\text{EC}}$.
\end{enumerate}
Every intermediate node introduced by changing an edge to a path increases $Q(R_\text{EC})$ by one compared to $Q(R_\text{BM})$. The naïve benchmark approach is therefore at least as good as when considering EC. To get an upper bound on $Q(R_\text{EC})$, we use that the worst case is when $\abs{\mathcal{T}}\geq N-1$ such that all users may request EPR pairs with all other users among the tasks, and $D$ is chosen such that the users can only be directly entangled with at most two other users. This corresponds to the case where $U_{i,j}^\mathcal{T}=1$ for all non-diagonal entries and the users can be represented in a sequence. The additional qubits required by introducing EC can then be calculated as the number of intermediate users in all possible paths in the sequence. Hence, we obtain that
\begin{align}
    Q(R_\text{EC})=Q(R_\text{BM})+\sum_{i=1}^{N-2}i(N-1-i)=\order{N^3}.
\end{align}
By introducing EC, the required number of qubits in the resource state has a cubic growth with the number of network users rather than a quadratic growth. Remark that this a loose upper bound that may be tightened by taking constraints on $\abs{\mathcal{T}}$ into account.

Notice that SED and EC can be considered simultaneously by changing the condition in step 1) to $U^{\mathcal{T}_\text{SED}}=1$ such that an edge has been removed in each task prior to considering EC. We denote the resource state of this approach $R_\text{SED + EC}$. By combining the arguments from Sec.~\ref{sec: Optimizing_SED} and Sec.~\ref{sec: Optimizing_EC}, we get that $Q(R_\text{SED + EC})=\order{N^3}$. In the case where the EPR pairs to be distributed by the satellite is such that no task has any remaining EPR pair of length greater than $D$, then $Q(R_\text{SED + EC})=Q(R_\text{SED})$.

\begin{table}[]
    \centering
    \begin{tabular}{c|c|c}
        Resource state & $R_\text{BM}, R_\text{SED}$ & $R_\text{EC}, R_\text{SED + EC}$ \\
        \hline
        \\[-1em]
        Computational complexity & $\order{N^2\abs{\mathcal{T}}}$ & $\order{N^3\abs{\mathcal{T}}}$ \\
        Initial size of resource state & $\order{N^2}$ & $\order{N^3}$ \\
    \end{tabular}
    \caption{Computational complexity of determining $U^{\mathcal{T}_{(\cdot)}}$ and worst-case order of $Q(R_{(\cdot)})$ for the four different settings for a network of $N$ users and $\abs{\mathcal{T}}$ possible tasks with $\abs{\mathcal{T}}\geq N-1$.}
    \label{tab: Algorithm_Complexity}
\end{table}

\subsection{The merging algorithm}
After determining one of the four resource states, we enhance its performance by using it as input for the merging algorithm proposed in \cite{b1}. Originally, the input to this algorithm is $\mathcal{T}$, which is used to calculate $R_\text{BM}$ as an initial resource state. However, we have designed $R_\text{SED}, R_\text{EC}$, and $R_\text{SED + EC}$ such that the algorithm also can handle these as initial resource states in our setting. The idea of the merging algorithm is for all users to test whether any pair of its qubits in the initial resource state can be merged into one with the merging measurement while still yielding a successful network. For a pair of qubits $i_1,i_2$ at user $i$, this is tested with the following two criteria in the respective order:
\begin{enumerate}
    \item If the incident edges of qubits $i_1$ and $i_2$ all have unit entries in $S^\mathcal{T}$, the two qubits are merged.
    \item Compute the reduced simultaneous redundancy matrix, $\widetilde{S}^\mathcal{T}$, which is obtained by only considering the first and second neighborhoods of qubits $i_1$ and $i_2$. That is, $\widetilde{S}^\mathcal{T}$ is computed by only considering the edges incident to all qubits in $N(i_1)$ and $N(i_2)$. If all incident edges of $i_1$ and $i_2$ have unit entries in $\widetilde{S}^\mathcal{T}$, the two qubits are merged.
\end{enumerate}
The output of the merging algorithm is an optimized resource state compared to the initial resource state. If a setting different from the benchmark setting is considered, the simultaneous adjacency matrix is computed based on another set of tasks, e.g. $S^{\mathcal{T}_\text{SED}}$. The most expensive part of the merging algorithm is determining the edges used to compute $\widetilde{S}^\mathcal{T}$ with cost $\order{Q(R_{(\cdot)})^2}$, which must be repeated for $\order{N[Q(R_{(\cdot)})/N]^2}$ pairs of qubits, yielding a total computational complexity of $\order{Q(R_{(\cdot)})^4/N}$. From this and Table~\ref{tab: Algorithm_Complexity}, it follows that the algorithm has a polynomial complexity for all settings.

\section{Results}\label{sec: Results}
In this section, we compare the performance of networks under the different settings for different parameters. For all simulations, the considered terrestrial network is a $5\times5$ grid similar to Fig.~\ref{fig: Physical_Network_with_Satellite} with $N$ users. In the cases where EC is considered, we use $D=2$. There are $\abs{\mathcal{T}}$ tasks for each network, and the number of EPR pairs in each task is randomly drawn from a Binomial distribution with $\lfloor N/2 \rfloor$ trials and success probability $p=0.8$. The users of the EPR pairs are chosen at random with the constraint that each user can occur in at most one EPR pair in each task. The four different settings we consider correspond to using the four different resource states presented in Sec.~\ref{sec: Optimizing_Resource} as input to the merging algorithm:
\begin{enumerate}
    \item The original setting without SED or EC ($R_\text{BM}$)
    \item Only EC ($R_\text{EC}$)
    \item Only SED ($R_\text{SED}$)
    \item Both SED and EC ($R_\text{SED + EC}$).
\end{enumerate}
To distinguish the settings in the plots with results, we use lines and dots to indicate whether EC is considered or not, respectively. Similarly, red and black colors distinguish whether SED is considered or not, respectively.

We use the physical network in Fig.~\ref{fig: Physical_Network_with_Satellite}, i.e., $N=6$ users, and $\abs{\mathcal{T}}=N-1$ as a starting point. We then vary different parameters to see their effect on the performance.

\begin{figure*}
    \centering
    \hspace*{5mm}
    \subfloat{%
        \includegraphics[width=0.9\linewidth]{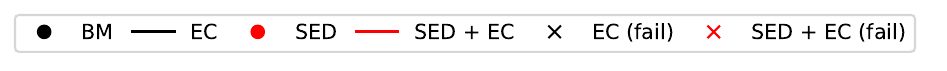}}\\
    \vspace*{-5mm}
    \setcounter{subfigure}{0}
    \subfloat[Different sets of tasks.\label{Results_NEW_Different_Tasks}]{%
        \includegraphics[width=.45\linewidth]{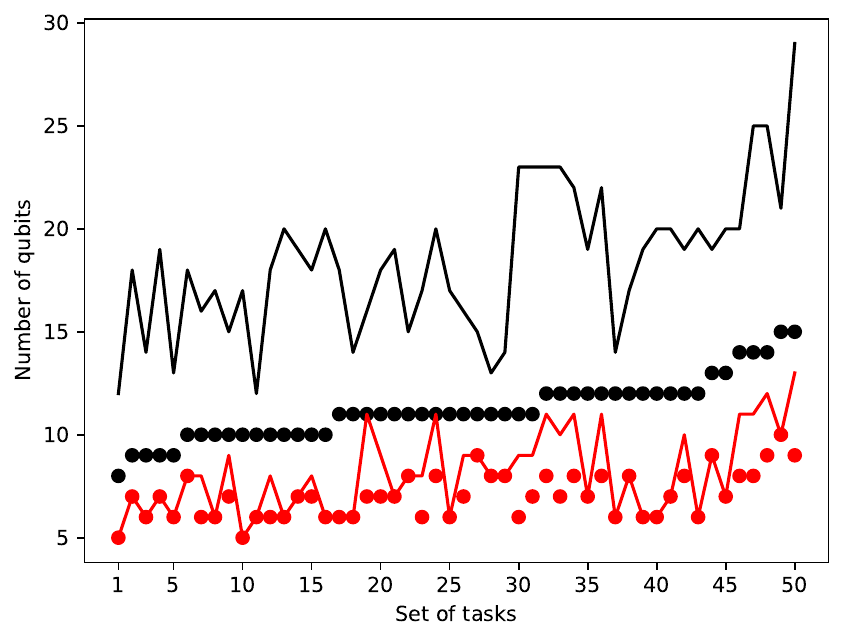}}
    \subfloat[Different user positions.\label{Results_NEW_Different_Positions}]{%
        \includegraphics[width=.45\linewidth]{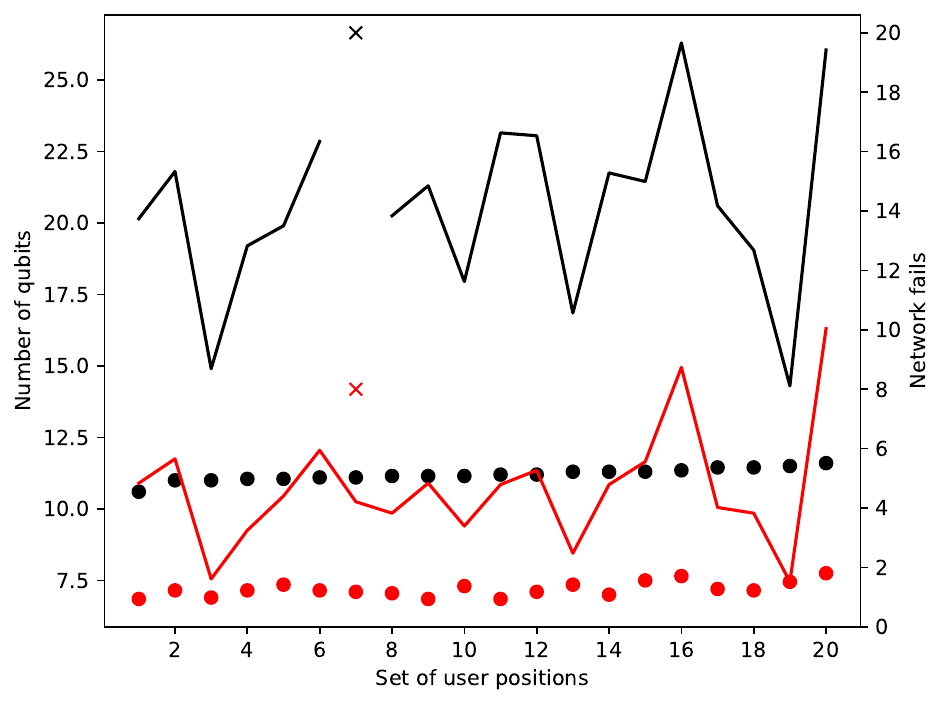}}\\
    \subfloat[Different number of users.\label{Results_NEW_N}]{%
        \includegraphics[width=.45\linewidth]{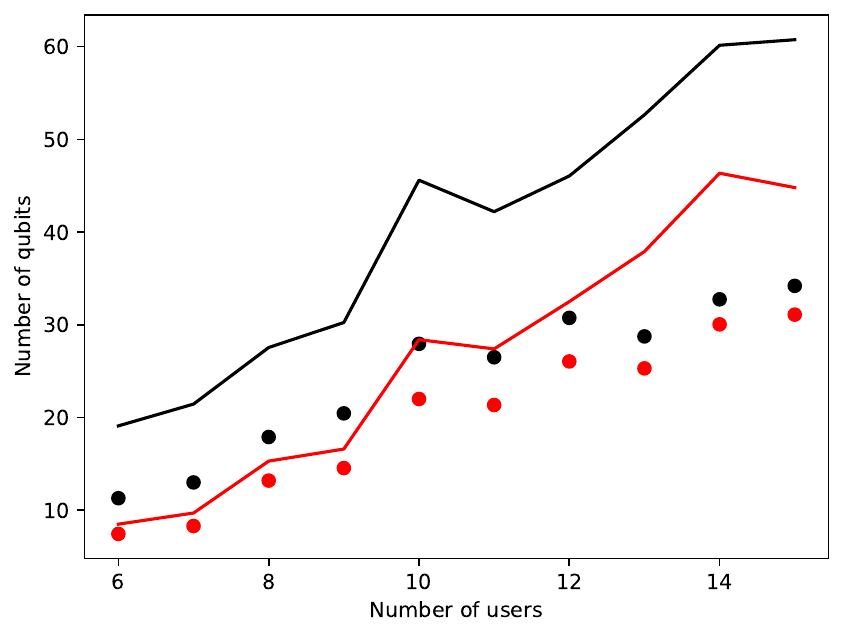}}
    \subfloat[Different sizes of set of tasks.\label{Results_NEW_M}]{%
        \includegraphics[width=.45\linewidth]{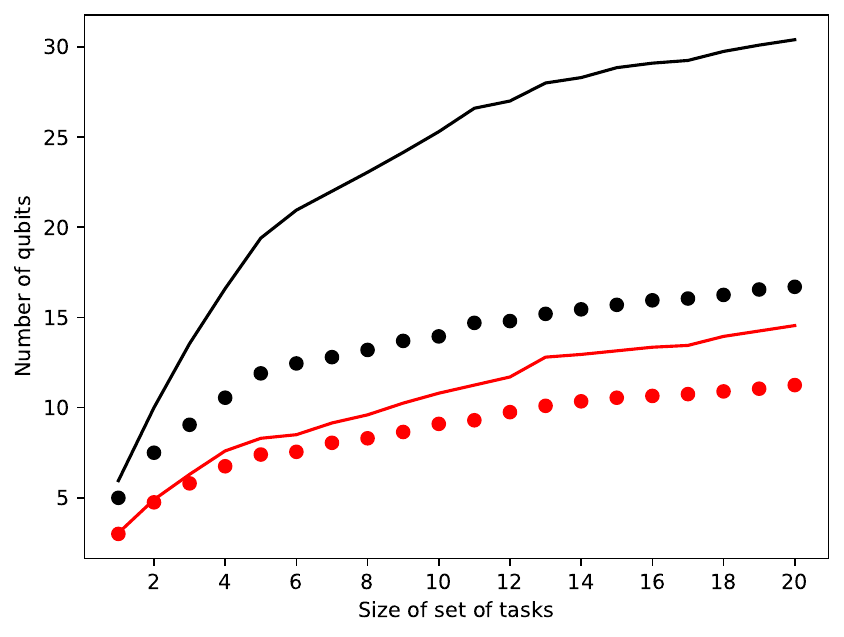}}
    \caption{Performance of the network for the four settings for different scenarios. A physical network topology with $N=6$ users positioned as in Fig.~\ref{fig: Physical_Network_with_Satellite} and $\abs{\mathcal{T}}=N-1$ is used as the starting point. (a) Different sets of tasks, $\mathcal{T}$, are drawn. (b) Average of $20$ sets of tasks for different user positions in the physical network topology. (c) Average of $20$ sets of tasks for an increasing number of users. (d) Average of $20$ sets for varying $\mathcal{T}$.}
    \label{fig: Results_NEW}
\end{figure*}

In Fig.~\ref{Results_NEW_Different_Tasks}, we compare the performance of the network for different sets of tasks. The figure illustrates that taking EC into account generally increases the required resources in the resource state compared to the BM setting while utilizing SED reduces the resources as expected. More precisely, when compared to $R_\text{BM}$,  $R_\text{EC}$ requires up to $111\%$ additional qubits while $R_\text{SED}$ and $R_\text{SED + EC}$ require up to $50\%$ fewer qubits. Furthermore, we see that $R_\text{SED + EC}$ always has better or identical performance compared to $R_\text{BM}$, which indicates that the availability of SED has a greater impact on the performance of this particular network than the EC caused by the physical network topology.

In Fig.~\ref{Results_NEW_Different_Positions}, we let the position of the $N=6$ users vary. For each set of user positions, we find the average performance for $20$ different sets of tasks. As the set of user positions now may contain a user with distance greater than $D=2$ to all other users, we also count the number of times the network is unsuccessful on the right $y$-axis (only plotted if nonzero). The performance of the different settings show the same tendency as in Fig.~\ref{Results_NEW_Different_Tasks}, however, $R_\text{SED + EC}$ is not superior to $R_\text{BM}$ for all user positions. In fact, for a single set of user positions (network $7$), the network is unsuccessful for all tasks for $R_\text{EC}$ and eight times for $R_\text{SED + EC}$. We note that the failures in network $7$ for $R_\text{SED + EC}$ stem from the simplistic way that an EPR pair is removed from each task when considering SED, namely that if multiple longest EPR pairs exist, we simply pick one of those at random. By chance, this may sometimes remove the `wrong' link, where another choice would have led to a successful network. We refrain from analyzing more intelligent choices in this work. Despite this simplistic approach, we still reduce the times of failure, which combined with the enhanced performance confirm the potential of SED in this framework.

In Fig.~\ref{Results_NEW_N}, we vary $N$ by gradually introducing a new user into the network topology. Again, we use the network topology in Fig.~\ref{fig: Physical_Network_with_Satellite} as starting point. By increasing $N$, we also increase the expected number of EPR pairs in each task as well as the size of the set of tasks by still using $\abs{\mathcal{T}}=N-1$. For each $N$, we compute the average performance for $20$ sets of tasks. As expected, the required resources increase when increasing the number of users. However, the increase when considering EC is more critical, particularly for $N\geq10$. From this point, $R_\text{SED + EC}$ no longer has superior performance compared to $R_\text{BM}$, indicating that EC as a result of the the physical network topology has greater impact on the performance of the network than the availability of SED.

Lastly, in Fig.~\ref{Results_NEW_M}, we once again fix the physical network topology to that in Fig.~\ref{fig: Physical_Network_with_Satellite}, i.e., $N=6$. Now, we vary the number of tasks $\abs{\mathcal{T}}$. The performance is again the average of $20$ sets of tasks for each value of $\abs{\mathcal{T}}$. All of the four settings can be seen to show an almost logarithmic growth as a result the fact that increasing $\abs{\mathcal{T}}$ generally also increase the amount of overlapping EPR pairs among the tasks due to only having $6$ users. However, the growth of $R_\text{EC}$ is extremely poor compared to the other settings as the probability of user $5$ requesting EPR pairs with all other users increases with $\abs{\mathcal{T}}$.

\section{Conclusion}\label{sec: Conclusion}
We have proposed a realistic yet idealized setting for the top-down approach for entanglement distribution in quantum networks involving two generalizations; 1) restrictions on the entanglement that can be created in the resource state due to the complexity of quantum routing, 2) enabling satellite-aided entanglement distribution in the adaptive phase to reduce decoherence in the static phase. We show how the merging algorithm used to optimize the number of qubits in the resource can be generalized to our proposed setting. Through numerical simulations, we find that satellite-aided entanglement distribution can enhance the performance of networks significantly while entanglement constraints has the opposite effect. This suggests that satellite-aided entanglement distribution to complement the resource state offers a promising direction for future research in the field. For small networks, the availability of the satellite generally has a greater impact on the performance than the physical network constraints, however, the opposite holds for larger networks. The problem with the poor scaling of large networks when considering entanglement constraints can be overcome by reducing the complexity of quantum routing through optimized routing protocols or optimized hardware such that less strict distance threshold can be used. Future work in this framework could explore a less idealized setting to better reflect practical applications or optimization of the algorithms for determining resource states.

\section*{Acknowledgment}
The authors acknowledge that the research was by the Villum Investigator
Grant “WATER” from the Velux Foundations, Denmark, and Centers of Excellence Grant ``CLASSIQUE'' from Denmark's Basic Research Foundation.

\end{document}

%% file: Preamble.tex
\usepackage{cite}
\usepackage{amsmath,amssymb,amsfonts}
\usepackage{algorithmic}
\usepackage{graphicx, ifthen}
\usepackage{subfig}
\usepackage{tikz, pgfplots}
\usetikzlibrary{positioning, fit, calc}
\usepackage{textcomp}
\usepackage{hyperref}
\usepackage{enumerate}
\usepackage{xcolor}
\usepackage{braket}
\usepackage{physics}
\usepackage{comment}

\def\BibTeX{{\rm B\kern-.05em{\sc i\kern-.025em b}\kern-.08em
    T\kern-.1667em\lower.7ex\hbox{E}\kern-.125emX}}

%% file: Figures/Phases_Time_Diagram2.tex
\begin{tikzpicture}[scale=.95, transform shape]
        
    \begin{axis}[
        width = .9\linewidth,
        height= .5\linewidth,
        xmin =0, xmax=9.5,
        ymin =0, ymax=1.4,
        clip=false,
        axis lines = middle,
        xtick = {2.5,4.7,8.6},
        xticklabels = {$t_1$, $t_2$, $t_3$},
        ytick = {.25,.5,.75,1},
        yticklabels = {Initialization, Dynamical, Static, Adaptive},
        ticklabel style = {font=\small},
        xlabel style = {at={(axis description cs:1.07,0)},anchor=north east},
        ylabel style = {at={(axis description cs:0,.9)},anchor=south east},
        xlabel = {\footnotesize{Time}},
        ylabel = {\footnotesize{Phase}},
        ]
        \addplot[domain=0:.5,black, line width=1pt]{.25};
        \addplot[domain=.5:1.5,black, line width=1pt]{.5};
        \addplot[domain=1.5:2.5,black, line width=1pt]{.75};
        \addplot[domain=2.5:3,black, line width=1pt]{1};
        \addplot[domain=3:3.7,black, line width=1pt]{.5};
        \addplot[domain=3.7:4.7,black, line width=1pt]{.75};
        \addplot[domain=4.7:5,black, line width=1pt]{1};
        \addplot[domain=5:5.6,black, line width=1pt]{.25};
        \addplot[domain=5.6:7.5,black, line width=1pt]{.5};
        \addplot[domain=7.5:8.6,black, line width=1pt]{.75};
        \addplot[domain=8.6:9,black, line width=1pt]{1};

        \addplot[thin, dashed, gray] coordinates {(2.5, 0) (2.5, 1)};
        \addplot[thin, dashed, gray] coordinates {(4.7, 0) (4.7, 1)};
        \addplot[thin, dashed, gray] coordinates {(8.6, 0) (8.6, 1)};

        \addplot[only marks, mark options={scale=.7}] coordinates {(.5,.25) (1.5,.5) (2.5,.75) (3,1) (3.7, .5) (4.7,.75) (5,1) (5.6,.25) (7.5,.5) (8.6,.75) (9,1)};
        \addplot[only marks, fill=white, mark options={scale=.7}] coordinates {(.5,.5) (1.5,.75) (2.5,1) (3,.5) (3.7,.75) (4.7,1) (5,.25) (5.6,.5) (7.5,.75) (8.6,1)};

        \addplot[thin, dashed, red] coordinates {(.5, -.3) (.5, 1)};
        \addplot[thin, dashed, red] coordinates {(3, -.3) (3, 1)};
        \addplot[<->, red] coordinates {(.5, -.3) (3, -.3)};
        \addplot[thin, dashed, red] coordinates {(5, -.3) (5, 1)};
        \addplot[<->, red] coordinates {(3, -.3) (5, -.3)};
        \addplot[thin, dashed, red] coordinates {(5.6, -.3) (5.6, 1)};
        \addplot[thin, dashed, red] coordinates {(9, -.3) (9, 1)};
        \addplot[<->, red] coordinates {(5.6, -.3) (9, -.3)};

        \node[] at (axis cs: 1.75,-.4) {\textcolor{red}{$\tau_1$}};
        \node[] at (axis cs: 4,-.4) {\textcolor{red}{$\tau_2$}};
        \node[] at (axis cs: 7.3,-.4) {\textcolor{red}{$\tau_3$}};

        \node[above right] at (axis cs:1.4,1) {\includegraphics[width=1cm]{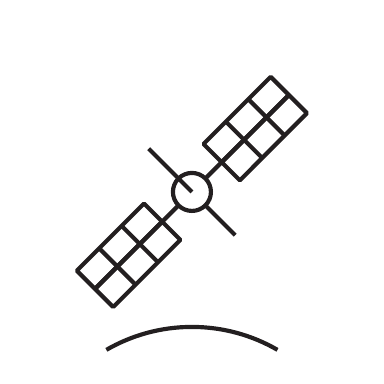}};
        \node[above right] at (axis cs:3.6,1) {\includegraphics[width=1cm]{Figures/Objects_satellite_.pdf}};
        \node[above right] at (axis cs:7.5,1) {\includegraphics[width=1cm]{Figures/Objects_satellite_.pdf}};
        
    \end{axis}
    
\end{tikzpicture}

%% file: Figures/Physical_Network_with_Satellite.tex
\tikzset{user/.style={rectangle, draw, red, fill=white, text=black, font=\huge, inner sep=3pt},
repeater/.style={circle, draw, gray, fill=gray, very thin, inner sep=2pt}
}

\makeatletter
\long\def\ifnodedefined#1#2#3{%
    \@ifundefined{pgf@sh@ns@#1}{#3}{#2}%
}
\makeatother

\makeatletter
\newcommand{\Letter}[1]{\@Alph{#1}}
\makeatother

\begin{tikzpicture}[scale=.5, transform shape]
    \filldraw[blue!40!black, fill=blue!10] ([shift={(11,-1)}]135:1) arc[radius=10, start angle=30, end angle=150];
    
    \filldraw[color=orange!60, fill=orange!5, very thick] (2,2) circle (3);
    
    \draw[blue!80] (0,0) grid (4,4);

    \node[user] (3) at (3,0) {6};
    \node[user] (4) at (4,0) {5};
    \node[user] (7) at (2,1) {4};
    \node[user] (9) at (4,1) {3};
    \node[user] (12) at (2,2) {1};
    \node[user] (15) at (0,3) {2};

    \foreach \j in {0,...,24}
    {
    \pgfmathsetmacro{\devicex}{mod(\j,5)}
    \pgfmathsetmacro{\devicey}{div(\j,5)}
    \ifnodedefined{\j}{}{
    \node[repeater] (\j) at (\devicex, \devicey) {};}
    }

    \node[] at (2,8) {\includegraphics[width=4cm]{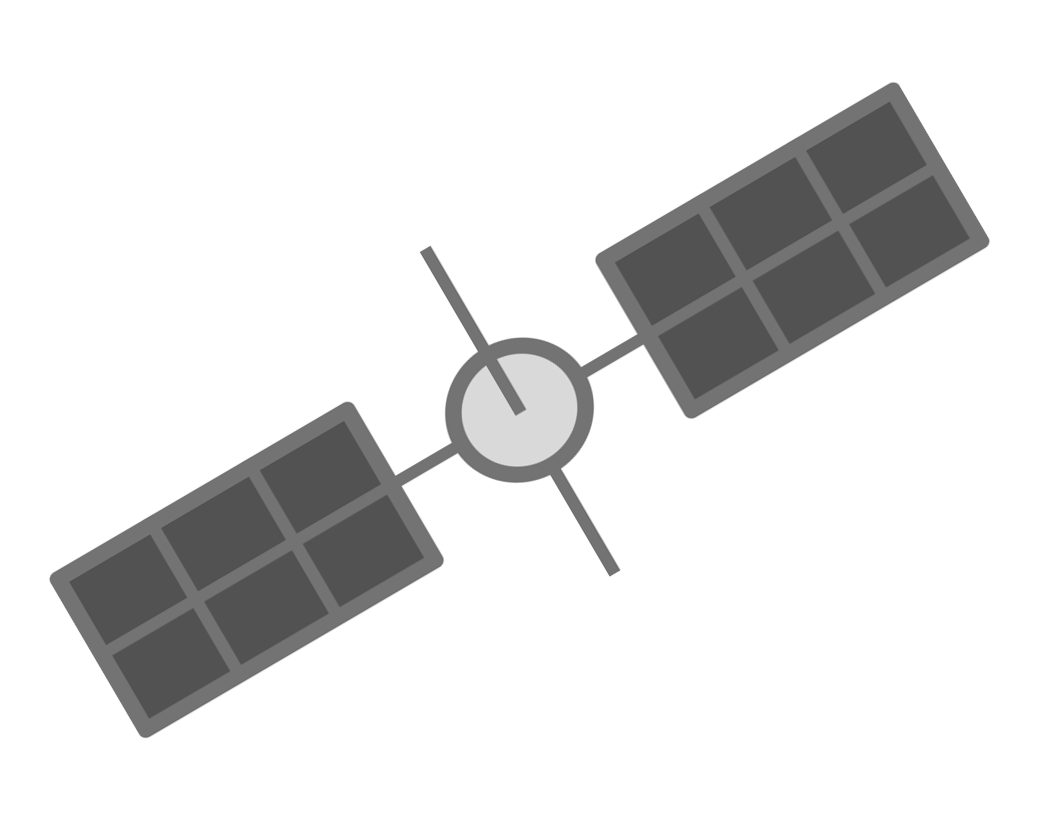}};
    \draw[color=orange!60] (2,7.72) -- (0,4.2);
    \draw[color=orange!60] (2,7.72) -- (4,4.2);
    
\end{tikzpicture}

%% file: Figures/Paper_Example_Resource.tex
\tikzset{network/.style={rectangle, draw, inner sep=8pt},
qubit/.style={circle, blue, draw, fill, very thin, inner sep=4pt},
user/.style={rectangle, draw, red, inner sep=6pt}
}

\begin{tikzpicture}[scale=.6, transform shape]

\node[qubit] (1A) {};
\node[user, fit=(1A)] (1) {};
\node[below right] at (1.north west) {1};

\node[qubit, right=2cm of 1A] (2A) {};
\node[user, fit=(2A)] (2) {};
\node[below right] at (2.north west) {2};

\node[qubit, right=2cm of 2A] (3A) {};
\node[qubit, right=.5cm of 3A] (3B) {};
\node[user, fit=(3A)(3B)] (3) {};
\node[below right] at (3.north west) {3};

\node[qubit, below=2cm of 1A] (4A) {};
\node[user, fit=(4A)] (4) {};
\node[above right] at (4.south west) {4};

\node[qubit, right=2cm of 4A] (5A) {};
\node[qubit, right=.5cm of 5A] (5B) {};
\node[user, fit=(5A)(5B)] (5) {};
\node[above right] at (5.south west) {5};

\node[qubit, right=2cm of 5B] (6A) {};
\node[user, fit=(6A)] (6) {};
\node[above right] at (6.south west) {6};

\draw[blue] (1A) -- (4A);
\draw[blue] (1A) -- (5A);
\draw[blue] (5A) -- (2A);
\draw[blue] (2A) -- (3A);
\draw[blue] (3B) -- (5B);
\draw[blue] (3B) -- (6A);

\node[network, fit=(1)(2)(3)(4)(5)(6)] (network) {};

\end{tikzpicture}

%% file: Figures/Paper_Example_Requets.tex
\tikzset{network/.style={rectangle, draw, inner sep=6pt},
qubit/.style={circle, blue, draw, fill, very thin, inner sep=2pt},
user/.style={rectangle, draw, red, inner sep=1.5pt}
}

\begin{tikzpicture}[scale=1, transform shape]

\node[qubit] (A1) {};
\node[qubit, right=10pt of A1] (A2) {};
\node[qubit, right=10pt of A2] (A3) {};
\node[qubit, white, below=10pt of A1] (A4) {};
\node[qubit, white, right=10pt of A4] (A5) {};
\node[qubit, right=10pt of A5] (A6) {};
\node[network, fit=(A1)(A2)(A3)(A6)] (1) {};

\draw[blue] (A1) -- (A2);
\draw[blue] (A3) -- (A6);

\foreach \i in {A1,A2,A3,A4,A5,A6}
{
\node[user, fit=(\i)] () {};
}

\node[qubit, right=50pt of A1] (B1) {};
\node[qubit, white, right=10pt of B1] (B2) {};
\node[qubit, white, right=10pt of B2] (B3) {};
\node[qubit, white, below=10pt of B1] (B4) {};
\node[qubit, right=10pt of B4] (B5) {};
\node[qubit, white, right=10pt of B5] (B6) {};
\node[network, fit=(B1)(B2)(B3)(B4)(B5)(B6)] (1) {};

\draw[blue] (B1) -- (B5);

\foreach \i in {B1,B2,B3,B4,B5,B6}
{
\node[user, fit=(\i)] () {};
}

\node[qubit, white, right=50pt of B1] (C1) {};
\node[qubit, right=10pt of C1] (C2) {};
\node[qubit, right=10pt of C2] (C3) {};
\node[qubit, below=10pt of C1] (C4) {};
\node[qubit, right=10pt of C4] (C5) {};
\node[qubit, white, right=10pt of C5] (C6) {};
\node[network, fit=(C1)(C2)(C3)(C4)(C5)(C6)] (1) {};

\draw[blue] (C2) -- (C4);
\draw[blue] (C3) -- (C5);

\foreach \i in {C1,C2,C3,C4,C5,C6}
{
\node[user, fit=(\i)] () {};
}

\node[qubit, right=50pt of C1] (D1) {};
\node[qubit, right=10pt of D1] (D2) {};
\node[qubit, right=10pt of D2] (D3) {};
\node[qubit, below=10pt of D1] (D4) {};
\node[qubit, right=10pt of D4] (D5) {};
\node[qubit, right=10pt of D5] (D6) {};
\node[network, fit=(D1)(D2)(D3)(D4)(D5)(D6)] (1) {};

\draw[blue] (D1) -- (D4);
\draw[blue] (D2) -- (D3);
\draw[blue] (D5) -- (D6);

\foreach \i in {D1,D2,D3,D4,D5,D6}
{
\node[user, fit=(\i)] () {};
}

\end{tikzpicture}

%% file: Figures/Example_Physical_Solution_D=5.tex
\tikzset{network/.style={rectangle, draw, inner sep=10pt},
qubit/.style={circle, blue, draw, fill, very thin, inner sep=4pt},
user/.style={rectangle, draw, red, inner sep=6pt}
}

\begin{tikzpicture}[scale=.4, transform shape]

\node[qubit] (1A) {};
\node[user, fit=(1A)] (1) {};
\node[below right] at (1.north west) {\Large 1};

\node[qubit, right=2cm of 1A] (2A) {};
\node[user, fit=(2A)] (2) {};
\node[below right] at (2.north west) {\Large 2};

\node[qubit, right=2cm of 2A] (3A) {};
\node[qubit, right=.5cm of 3A] (3B) {};
\node[user, fit=(3A)(3B)] (3) {};
\node[below right] at (3.north west) {\Large 3};

\node[qubit, below=2cm of 1A] (4A) {};
\node[user, fit=(4A)] (4) {};
\node[above right] at (4.south west) {\Large 4};

\node[qubit, right=2cm of 4A] (5A) {};
\node[qubit, right=.5cm of 5A] (5B) {};
\node[user, fit=(5A)(5B)] (5) {};
\node[above right] at (5.south west) {\Large 5};

\node[qubit, right=2cm of 5B] (6A) {};
\node[user, fit=(6A)] (6) {};
\node[above right] at (6.south west) {\Large 6};

\draw[blue] (1A) -- (4A);
\draw[blue] (1A) -- (5A);
\draw[blue] (2A) -- (3A);
\draw[blue] (3A) -- (5A);
\draw[blue] (3B) -- (5B);
\draw[blue] (3B) -- (6A);

\node[network, fit=(1)(2)(3)(4)(5)(6)] (network) {};

\end{tikzpicture}

%% file: Figures/Example_Physical_Solution_D=4.tex
\tikzset{network/.style={rectangle, draw, inner sep=10pt},
qubit/.style={circle, blue, draw, fill, very thin, inner sep=4pt},
user/.style={rectangle, draw, red, inner sep=6pt}
}

\begin{tikzpicture}[scale=.4, transform shape]

\node[qubit] (1A) {};
\node[qubit, right=.5cm of 1A] (1B) {};
\node[user, fit=(1A)(1B)] (1) {};
\node[below right] at (1.north west) {\Large 1};

\node[qubit, right=2cm of 1B] (2A) {};
\node[user, fit=(2A)] (2) {};
\node[below right] at (2.north west) {\Large 2};

\node[qubit, right=2cm of 2A] (3A) {};
\node[qubit, right=.5cm of 3A] (3B) {};
\node[qubit, right=.5cm of 3B] (3C) {};
\node[user, fit=(3A)(3B)(3C)] (3) {};
\node[below right] at (3.north west) {\Large 3};

\node[qubit, below=2cm of 1A] (4A) {};
\node[qubit, right=.5cm of 4A] (4B) {};
\node[user, fit=(4A)(4B)] (4) {};
\node[above right] at (4.south west) {\Large 4};

\node[qubit, right=2cm of 4B] (5A) {};
\node[qubit, right=.5cm of 5A] (5B) {};
\node[user, fit=(5A)(5B)] (5) {};
\node[above right] at (5.south west) {\Large 5};

\node[qubit, right=2cm of 5B] (6A) {};
\node[user, fit=(6A)] (6) {};
\node[above right] at (6.south west) {\Large 6};

\draw[blue] (1A) -- (4A);
\draw[blue] (1A) -- (5A);
\draw[blue] (1B) -- (2A);
\draw[blue] (1B) -- (4B);
\draw[blue] (3A) -- (4B);
\draw[blue] (3B) -- (5A);
\draw[blue] (3C) -- (5B);
\draw[blue] (3C) -- (6A);

\node[network, fit=(1)(2)(3)(4)(5)(6)] (network) {};

\end{tikzpicture}

%% file: Figures/Example_Solution3_Resource.tex
\tikzset{network/.style={rectangle, draw, inner sep=8pt},
qubit/.style={circle, blue, draw, fill, very thin, inner sep=4pt},
user/.style={rectangle, draw, red, inner sep=6pt}
}

\begin{tikzpicture}[scale=.6, transform shape]

\node[qubit] (1A) {};
\node[user, fit=(1A)] (1) {};
\node[below right] at (1.north west) {1};

\node[qubit, white, right=2cm of 1A] (2A) {};
\node[user, fit=(2A)] (2) {};
\node[below right] at (2.north west) {2};

\node[qubit, right=2cm of 2A] (3A) {};
\node[user, fit=(3A)] (3) {};
\node[below right] at (3.north west) {3};

\node[qubit, below=2cm of 1A] (4A) {};
\node[user, fit=(4A)] (4) {};
\node[above right] at (4.south west) {4};

\node[qubit, right=2cm of 4A] (5A) {};
\node[user, fit=(5A)] (5) {};
\node[above right] at (5.south west) {5};

\node[qubit, right=2cm of 5A] (6A) {};
\node[user, fit=(6A)] (6) {};
\node[above right] at (6.south west) {6};

\draw[blue] (1A) -- (4A);
\draw[blue] (3A) -- (5A);
\draw[blue] (5A) -- (6A);

\node[network, fit=(1)(2)(3)(4)(5)(6)] (network) {};

\end{tikzpicture}